\documentclass[12pt,a4paper]{article}
\usepackage[scale={.75,.8}]{geometry}
\usepackage{pstricks}
\usepackage{graphicx}
\usepackage[latin2]{inputenc}
\usepackage{epic}
\usepackage{latexsym}
\usepackage{epsf}
\usepackage{epsfig}
\usepackage{cite}
\usepackage{amssymb,amsmath}

\begin{document}

\vspace{-1truecm}

\rightline{CERN-PH-TH-2006-127}
\rightline{CPHT-RR 054.0706}
\rightline{LPT--Orsay 06/45}
\rightline{hep-th/0607077}
\rightline{July 2006}

\vspace{0.cm}

\begin{center}

{\Large {\bf Moduli stabilization with positive vacuum energy}}
\vspace{1 cm}\\

{\large E. Dudas$^{1,2,3}$ \ and \ Y. Mambrini$^3$
}
\vspace{1cm}\\

$^1$
CERN Theory Division, CH-1211, Geneva 23, Switzerland
\vspace{0.3cm}\\

$^2$
CPhT, Ecole Polytechnique 91128 Palaiseau Cedex, France
\vspace{0.3cm}\\

$^3$
Laboratoire de Physique Th\'eorique,
Universit\'e Paris-Sud, F-91405 Orsay, France
\vspace{0.3cm}\\

\end{center}

\vspace{1cm}

\abstract{
We study the effect of anomalous $U(1)$ gauge groups in
string theory compactification with fluxes. We find that,
in a gauge invariant formulation, consistent AdS vacua appear
breaking spontaneously supergravity. Non vanishing
D-terms from the anomalous symmetry act as an uplifting potential
and could allow for de Sitter vacua. However, we show that in this case the gravitino  is generically (but not
always) much heavier than the electroweak scale. We show that alternative uplifting scheme based on corrections to the 
Kahler potential can be compatible with a gravitino mass in the TeV range.}

\newpage

\vspace{3cm}

\newpage

\tableofcontents

\vspace{3cm}


\pagestyle{plain}

\section{Introduction and Conclusions}

Recently, Kachru et al. proposed an interesting
set--up to stabilize the moduli within the framework
of type IIB string theory orientifolds. The KKLT set--up \cite{Kachru:2003aw}
involves three steps to achieve a SUSY breaking Minkowski
vacuum, while stabilizing all moduli. The first step is to introduce
the NS and RR 3-form fluxes, $H_3$ and $F_3$, stabilizing the
dilaton $S$ and all complex structure moduli $U_{\alpha}$
\cite{Giddings:2001yu}.
In the second step, the overall Kahler modulus $T$ is fixed by non--perturbative
effects such as gaugino condensation \cite{Nilles:1982ik}.
The last step is to introduce an explicit SUSY breaking term
induced by anti $D3$-branes providing a positive uplifting potential
which would generate a dS vacuum. An obvious drawback of this procedure is the third step, the need of introducing
a sector with nonlinearly realized supersymmetry. It was later proposed \cite{Burgess:2003ic} to replace in the third step
the antibranes by a D-term spontaneous supersymmetry breaking sector induced by magnetic fluxes on the world-volume of
$D7$-branes. As shown later on \cite{Dudas:2005vv}, \cite{Villadoro:2005yq}, \cite{Achucarro:2006zf}, \cite{choi}
and reviewed in Section 2, however, gauge invariance imposes restrictions on the resulting
effective Lagrangian which makes this step difficult to reconcile with low-energy supersymmetry, namely D-terms uplifting
asks (in the absence of additional fine-tuning) for a very large gravitino mass, generically two orders of magnitude 
below the Planck mass. We show explicitly in Section 3 by analytical methods, in the case where the gaugino condensation 
scale is below the Fayet--Iliopoulos (FI) scale, that the generic value of the D-terms is of the order of 
D $\sim m_{3/2}^2$, 
whereas values needed for the uplifting are of order $D \sim m_{3/2} M_P$. It is possible to have much larger values
of the D-terms, of order $g^4 M_P^4$, where $g$ denotes some gauge coupling constant 
\cite{Villadoro:2005yq, Achucarro:2006zf},
at the price of  having no separation between the FI terms and the gaugino condensation scale, the outcome being 
a very large gravitino mass.   
Whereas this result can be avoided in more involved constructions, we believe that generically an additional fine-tuning
will be needed in order to keep the gravitino light.
   In the second part of this letter, we show that the situation changes significantly by including corrections to the
Kahler potential of the relevant modulus field $T$. Such corrections arise inevitably through quantum corrections and/or
$\alpha'$ corrections in string theory. We show in general and exemplify with an example that in this case the uplifting
is more likely to be compatible with low energy supersymmetry. We leave for future work a complete phenomenological 
analysis of the uplifting models we studied from the point of view of the computation of soft terms and their 
low-energy consequences.

The plan of this letter is as follows. Section 2 discusses the issue of D-term uplifting versus gauge invariance. 
In section 3 we work out analytically the example of a hidden sector coupled to an anomalous $U(1)$ gauge 
symmetry, in the limit
where the FI scale is much higher than the gaugino condensation scale, in both cases of positive and negative
value of the FI term. In the first case, the requirements of zero vacuum energy and TeV scale gravitino mass asks
for unrealistically low values of the FI term in the TeV range, whereas natural values of the FI term lead to 
gravitino masses larger than about $10^{-2} M_P$. In the second case, zero vacuum energy with TeV gravitino 
mass asks for intermediate
values of the FI term, whereas natural values of the FI term lead to gravitino masses larger than about $10^{-6} M_P$. 
  Section 4 presents the general features of Kahler uplifting and an explicit example
illustrates the compatibility between the Kahler uplifting and TeV values for the gravitino mass.   
   
\section{D-terms, gauge invariance and anomalies}
\subsection{The heterotic case}
Gaugino condensation in heterotic theories in the presence of the (generic) anomalous $U(1)$ gauge symmetry
has to fulfill the consistency requirements dictated by the
coexistence of the two local symmetries : supersymmetry and the
gauge symmetry. Indeed, $U(1)$ gauge transformations act as\footnote{We use here the same convention as in
\cite{Binetruy:1996uv} to define charges of chiral superfields.}
\begin{eqnarray}
&& \delta V_X \ = \  \Lambda \ + \ {\bar \Lambda} \qquad , \quad 
\delta \phi^i \ = \ - 2 \ q_i \ \phi^i \ \Lambda \ \equiv \ -2 X^i \Lambda \ , \nonumber \\
&& \delta S \ = \ \delta_{GS} \ \Lambda \ \equiv \ -2 \ X^S \Lambda \ ,   \label{gi1}
\end{eqnarray}
where $X^i,X^S$ define the holomorphic Killing vectors, which appear in the D-term 
\begin{equation}
D_X \ = \ X^i \ \partial_i G \ = \  X^i \ \partial_i K \ = \    q_i \ \phi^i \ \partial_i K \ + 
\ {\delta_{GS} \over 2 (S + {\bar S})} \ , \label{gi01} 
\end{equation}
where in (\ref{gi01}) $G = K + \ln |W|^2$ and we used the gauge invariance of the superpotential $X^i \partial_i W = 0$. 
The FI term is encoded in the modified Kahler potential for the universal dilaton-axion
\begin{equation}
K \ = \ - \ \ln \ (S \ + \ {\bar S} \ - \ \delta_{GS} V_X) \ . \label{gi02} 
\end{equation}
 The pure Super--Yang--Mills gaugino condensation 
superpotential $e^{-3 S/2b_0 }$, where $b_0$ is the beta function of the hidden sector, is therefore
not gauge invariant. Does this mean that gauge
invariance forbids gaugino condensation to take place ? The answer
to this question in the simpler situation of the heterotic string
was given some time ago in \cite{Binetruy:1996uv}. It was shown
there that the Green-Schwarz (GS) cancellation of gauge anomalies
restricts the nonperturbative dynamics such that the nonperturbative
superpotential is precisely gauge invariant. We summarize here the
argument in order to generalize it later on to orientifold models with open string
fluxes. In the perturbative heterotic constructions, there is only
one possible anomalous $U(1)_X$ and one field, the universal
axion-dilaton $S$, transforming non-linearly under gauge
transformations (\ref{gi1}). Anomaly cancellation
conditions relate mixed anomalies $C_i = U(1)_X G_i^2$, where $G_i$
are the various semi-simple factors of the gauge group $G =
\prod_{i=1}^N G_i$, such that

\begin{equation}
\delta_{GS} \ = \ {C_1 \over k_1} \ =  \ {C_2 \over k_2} = \cdots {C_N
\over k_N} \ = \ {1 \over 192 \pi^2} Tr q \ , \label{gi2}
\end{equation}
where $k_i$ are the Kac-Moody levels defining the tree-level gauge
kinetic functions
\begin{equation}
f_i \ = \ k_i \ S \ . \label{gi3}
\end{equation}
The last term in (\ref{gi2}) is the Fayet-Iliopoulos (FI) term, proportional to the mixed $U(1)_X$ - gravitational
anomaly, where $Tr q$ is the sum of $U(1)$ charges over all the charged
fermions in the spectrum. Therefore, once the FI term is generated,
all mixed anomalies have to be different from zero and the hidden
sector {\it must} contain charged matter. Taking for simplicity a
SUSY--QCD with $N_c$ colors and $N_f$ flavors with $N_f < N_c$ and
denoting by $Q$ ($\tilde Q$) the hidden sector quarks (antiquarks)
of $U(1)$ charges $q$ (${\tilde q}$), the GS conditions fix
completely the sum of the charges to be
\begin{equation}
C_h \ = \ {1 \over 4 \pi^2} \ N_f (q + {\tilde q}) \ = \ \delta_{GS} k_h \ . \label{gi4}
\end{equation}
This turns out to be precisely the gauge invariance condition of the
nonperturbative superpotential
\begin{equation}
W_{np} \ = \ (N_c-N_f) \left[ {e^{- 8 \pi^2 k_h S} \over det (Q {\tilde Q})}
\right]^{1 \over N_c-N_f} \ . \label{gi5}
\end{equation}
Notice that anomaly cancellations (\ref{gi2}) and the structure of the D-term (\ref{gi01}) unambigously shows that the the charge 
of the hidden sector mesons $Q {\tilde Q}$ has the same sign as the induced FI term. Since in all heterotic models there is at least
one scalar $\phi$ of the appropriate (negative, in our conventions in what follows) charge in order to be able to compensate the FI term at tree-level,
the simplest model adressing moduli stabilisation with D-terms contain the modulus $S$, the hidden sector meson fields and $\phi$. 

\subsection{Orientifolds with internal magnetic fields}

In order to fix $T$, KKLT proposes to use background fluxes
for both NS and RR forms to fix the complex--structure moduli
to obtain a vacuum in which supersymmetry is broken by the $T$
field because $W=W_0 \neq 0$. Then they introduce a
non--perturbative superpotential of the form $W_{np}=ce^{-aT}$
induced by gaugino condensations on $D7$-branes or 
$D3$-branes euclidian instantons. Combining the two sources of superpotentials after having
integrated out all complex--structure and dilaton fields, they considered\footnote{See e.g. \cite{gauginofluxes} for more detailed
analysis of nonperturbative superpotentials in orientifold models.}

\begin{equation}
W \ = \ W_0 \ + \ c \ e^{-aT} \ . \label{WKKLT}
\end{equation}

\noindent The resulting vacuum is supersymmetric Anti de Sitter. To
obtain a phenomenological desirable de Sitter or Minkowski vacuum,
the authors then proposed to uplift the energy with a positive
contribution to the potential from anti $D3$-branes. This has the
effect of adding an extra non-supersymmetric contribution to the
scalar potential of the form :

\begin{equation}
V \ = \ V_F \ + \ \frac{k}{T_R^2} \ , 
\label{VKKLT}
\end{equation}

\noindent where $k$ is a (fine--tuned) constant and $T_R =
\mathrm{Re} \ T$. For a suitable value of $k$, the original AdS vacuum
get lifted to a dS one with broken (nonlinearly realized)
supersymmetry.

Later on Burgess et al. \cite{Burgess:2003ic} proposed an alternative solution by replacing the anti $D3$-brane
contribution with a $D-$term contribution originated from
magnetic fluxes. These fluxes would in turn generate a
Fayet--Iliopoulos (FI) term in the 4D effective action of the form

\begin{equation}
V_D \ = \ {g^2 \over 2} \ D^2 \ = \ \frac{2 \pi}{T_R} \left( \sum_i
q_i K_i \phi^i + \xi \right)^2  \ , \label{VDquevedo}
\end{equation}

\noindent where the $\phi^i$ represent any matter scalar fields
which are charged (with charge $q_i$) under the $U(1)_X$ gauge
group. $K_i$ is the derivative of the Kahler potential $K$ with
respect to the field $\phi^i$, and the FI term $\xi$ arises from
non--trivial fluxes for the gauge fields living on the $D7$-branes
and is given by

\begin{equation}
\xi \ = \ - \ {\delta_{GS} \over 2} \ \partial_T K \ .
\end{equation}

However, as was stressed in \cite{Dudas:2005vv},
\cite{Villadoro:2005yq} and \cite{Achucarro:2006zf} such proposal
does not fulfill the consistency requirements dictated by the
coexistence of the two local symmetries : supersymmetry and the
gauge symmetry. Indeed, $U(1)$ gauge transformations act as

\begin{eqnarray}
&& \delta V \ = \  \Lambda + {\bar \Lambda} \qquad , \quad \delta \phi^i \ = \ -2 \ q_i \phi^i \Lambda \ , \nonumber \\
&& \delta T \ = \ \delta_{GS} \Lambda \  \label{gi6}
\end{eqnarray}
and, in analogy with the heterotic case discussed earlier, the pure Super--Yang--Mills gaugino condensation superpotential
$e^{-a T }$ is not gauge invariant. 

In the case of orientifold models with magnetic fluxes in the compact space, the fields playing a role
in the anomaly cancellation are Kahler moduli $T_i$, whereas in the intersecting branes language they are
the complex structure moduli $U_i$. Several such fields can participate in canceling triangle anomalies
and therefore anomaly cancellation conditions are generically more complicated than the heterotic universal
relation (\ref{gi2}). In particular, whereas in the heterotic case the $U(1)_X$ charge of the hidden sector mesons
$Q {\tilde Q}$ is always of the same sign as the induced FI term due to (\ref{gi2}), this is not necessarily
always the case in orientifold examples. We will use this possible difference later on in searching for
different uplifting possibilities.   

The appropriate context are orientifold models with open string magnetic fluxes \cite{magnetic}, T-dual (in the absence 
of closed string fluxes) 
to intersecting brane constructions \cite{intersecting}. For concreteness, we summarize the case of magnetized D9 branes, 
but the results
are actually more general\footnote{See \cite{vz2} for a more general discussion in orientifold models with fluxes.}. 
On the wordvolume of $D9^{(a)}$ branes we can introduce magnetic fields $H_1^{(a)},H_2^{(a)},H_3^{(a)}$ in the three torii 
describing
the three complex internal spaces, of volumes $v_1,v_2,v_3$. The generic gauge group coming from D9 branes is $\prod_a U( N_a) $, where $N_a$ is the
number of the branes in a given stack.  
The Dirac quantization condition in this case take the generalized form
\begin{equation}
H_i^{(a)} \ = \ {m_i^{(a)} \over n_i^{(a)} v_i} \ , \label{gi7}
\end{equation}
where $(m_i^{(a)},n_i^{(a)})$ are integers.
Their interpretation is transparent in the T-dual D6 intersecting brane language, where the 
magnetic fields are mapped
into angles that the D6 branes have with one of each cordinates of the three torii
\begin{equation}
\tan \theta_i^{(a)} \ = \ H_i^{(a)} \ = \ {m_i^{(a)} R_{i2} \over n_i^{(a)} R_{i1}} \  \label{gi8}  
\end{equation} 
for rectangular torii of volumes $v_i = R_{i1} R_{i2}$. 
In this language,  $(m_i^{(a)},n_i^{(a)})$ are the wrapping numbers of the brane $D6^{(a)}$ along the 
two compact directions of the torus $T^2_i$.
The Born-Infeld and the Wess-Zumino couplings are
\begin{eqnarray}
&& {1 \over g_{(a)}^2} \ = \ | n_1^{(a)} n_2^{(a)} n_3^{(a)} | v_1 v_2 v_3 \prod_{i=1}^3 \sqrt{ 1 + H_i^{(a) 2} } \ , \nonumber \\
&& S_{WZ} \ = \   n_1^{(a)} n_2^{(a)} n_3^{(a)} \ \int C \ tr ( \ e^{ s_1 H_1 + s_2 H_2 + s_3 H_3} \ e^F ) \ , \label{gi9} 
\end{eqnarray}
where in the second line we used the form language, $s_i = \pm 1/2$ are internal fermionic helicities and where $F$ is the D9 brane gauge 
field strength. After compactification to
4d we can then write down the gauge couplings and the axionic couplings. From the axionic couplings and by keeping the three 
Kahler moduli $T_i$ present
in several 4d compactifications (for example, the $Z_2 \times Z_2'$ orbifold), we can guess the gauge kinetic function
\begin{equation}
f_a \ = \  n_1^{(a)} n_2^{(a)} n_3^{(a)} S \ - \  n_1^{(a)} m_2^{(a)} m_3^{(a)} T_1 \ - \  
m_1^{(a)} n_2^{(a)} m_3^{(a)} T_2
\ - \  m_1^{(a)} m_2^{(a)} n_3^{(a)} T_3 \ . \label{gi10}
\end{equation} 
The first remark is that the gauge functions in orientifolds with intersecting branes are highly dependent on how branes wrap
the compact space, to be compared to the universal form in the heterotic case (\ref{gi3}). 
Whereas $Im f$ in (\ref{gi10}) comes precisely from the Wess-Zumino coupling in (\ref{gi9}), $Re f$ in (\ref{gi10}) 
only matches the compactified gauge coupling in (\ref{gi9}) in the case
\begin{equation}
\theta_1^{(a)} \ + \  \theta_2^{(a)} \ + \  \theta_3^{(a)} \ = \ 0   \quad \leftrightarrow \quad  
H_1^{(a)} \ + \  H_2^{(a)} \ + \  H_3^{(a)} \ = \ H_1^{(a)} H_2^{(a)} H_3^{(a)}
\ , \label{gi11}
\end{equation}
which also define the condition for preserving supersymmetry in internal magnetic fields / intersecting 
branes constructions.  
In case where (\ref{gi11}) is violated, the gauge couplings and axionic couplings from the Born-Infeld 
action/ Wess-Zumino couplings do not
fit into the holomorphic $f_a$ gauge kinetic function. In addition,  the $U(1)^{(a)}$ part of the 
$U(N_a)$ gauge group becomes ``anomalous'', in the sense of acquiring Fayet-Iliopoulos terms
\begin{eqnarray}
&& \xi_a \ = \ {m_1^{(a)} n_2^{(a)} n_3^{(a)} \over Re \ T_1 } + {n_1^{(a)} m_2^{(a)} n_3^{(a)} \over Re \ T_2 } + 
{n_1^{(a)} n_2^{(a)} m_3^{(a)} \over Re \ T_3 } - {m_1^{(a)} m_2^{(a)} m_3^{(a)} \over Re \ S } \ \nonumber \\
&& \ \sim \ H_1^{(1)} \ + \ H_2^{(a)} \ + \ H_3^{(a)} \ - \ H_1^{(a)} \ H_2^{(a)} \ H_3^{(a)} \ . \label{gi12}  
\end{eqnarray}
Effective supergravity description  seems to be valid only in the limiting case $\xi_a \ll 1$, which has to be dynamically obtained
in models with moduli stabilisation. In this case, the FI terms can be incorporated into the effective action
via the modification of the Kahler potential
\begin{eqnarray}
&& K \ = \ - \ln \ (S + {\bar S} + m_1^{(a)} m_2^{(a)} m_3^{(a)} V_a ) \ - \ 
\ln \ (T_1 + {\bar T}_1 - m_1^{(a)} n_2^{(a)} n_3^{(a)} V_a ) \nonumber \\
&& - \ln \ (T_2 + {\bar T}_2 - n_1^{(a)} m_2^{(a)} n_3^{(a)} V_a ) \ - \ 
\ln \ (T_3 + {\bar T}_3 - n_1^{(a)} n_2^{(a)} m_3^{(a)} V_a ) \ , \label{gi13}
\end{eqnarray}
which also fixes the $U(1)_a$ gauge transformations of the moduli fields
\begin{eqnarray}
&& \delta V_a \ = \ \Lambda_a \ + \ {\bar \Lambda}_a \quad , \quad \delta S \ = \ -  m_1^{(a)} m_2^{(a)} m_3^{(a)} 
\ \Lambda_a
\ , \nonumber \\
&& \delta T_1 \ = \   m_1^{(a)} n_2^{(a)} n_3^{(a)} \ \Lambda_a \quad , \quad 
\delta T_2 \ = \   n_1^{(a)} m_2^{(a)} n_3^{(a)} \ \Lambda_a \ , \nonumber \\
&& \delta T_3 \ = \   n_1^{(a)} n_2^{(a)} m_3^{(a)} \ \Lambda_a \ . \label{gi14}
\end{eqnarray}

Anomaly cancellation in orientifold models with fluxes is therefore considerably more involved than in the
heterotic case (not to mention the fact that twisted closed fields can and typically do also play a nontrivial
role in gauge and gravitational anomaly cancellation). 
A notable difference, which we will invoke later on, is that there is no clear cut correlation between the
sign of FI terms and charges of hidden (or observable) sector fields. 
Indeed, the induced FI terms (\ref{gi12}) can have either sign depending on the v.e.v.'s of moduli fields, whereas 
charges in a given sector depend on details of the model. Notice that it is also possible (but not automatic) by partial
cancellations between different contributions that $\xi_a \ll 1$, such that an effective supergravity analysis to be
valid. 

\section{D--terms uplifting}
\subsection{Stabilization with negative vacuum energy }

Recently, the authors of \cite{Villadoro:2005yq} and
\cite{Achucarro:2006zf} tried to implement gauge invariance
in the search for consistent de Sitter vacua. This was indeed
realized, at the expense of having a high scale of gaugino
condensation, close to the Planck scale. Consequently, these models
have a large gravitino mass, more appropriate for the split
supersymmetry scenario. It was concluded that it seems very hard to
obtain a gravitino mass in the TeV scale region. In the rest of this
section, we analyze, for a model with just one (overall) volume modulus $T$,  the difficulties arising from a consistent
D-term uplifting of the vacuum energy with TeV scale gravitino mass
and possible ways out. \footnote{During the completion of this work,
reference \cite{choi} appeared, which overlaps with our results in
Sections 3.1 and 3.2.}
The minimal model we consider here is that of a hidden sector coming from 
$D7$-branes, which is a SQCD with less number of
flavors $N_f $ than colors $N_c$. For simplicity of analysis we take $N_f=1$, $N_c=2$. The example assumes 
(this assumption, as we already explained, is always true in the heterotic case) that the FI term has the 
same sign (chosen positive, 
by convention) as the hidden sector meson charge. An important difference compared to the examples  in 
\cite{Villadoro:2005yq} and \cite{Achucarro:2006zf} is that we use the existing  string ``data'' in that, {\it in all
existing constructions, there is always at least one field of appropriate charge which, in the absence of nontrivial
superpotential dynamics, is able to compensate the FI term}. We call this field $\phi$ in what follows and assign it
the charge $-1$, after an appropriate rescaling of the charge generator. If the meson charge $q$ is integer, this will
have the consequence of inducing higher dimensional operators of the type $ \phi^q M $ which, once $\phi$ gets a vev,
effectively give the hidden sector mesons a mass. Since the gaugino condensation scale and consequently the
meson vev's $M_0$ arise through dimensional transmutation, they are expected to be well below the scale of the FI terms
$\xi$. This will allows us, following \cite{Binetruy:1996uv}, to solve analytically for the vev's of the various fields
and the scale of supersymmetry breaking, in an expansion in the small parameter $\epsilon = M_0 / \xi$.        

Following \cite{Binetruy:1996uv}, we write down the consistent effective
potential of a strongly coupled $SU(2)$ theory with $N_f=1$ flavor
of "quarks" $Q_1$ of $U(1)_X$ charge $q_1$ in the fundamental
of $SU(2)$ and "antiquarks" $\tilde{Q_1}$ of charge
$\tilde{q_1}$ in the anti-fundamental of $SU(2)$.
For simplicity in the calculation and analysis, from
the meson field $M$, we will define
the field $\chi$ as $\chi^2 = 2 M = 2 Q_1 \tilde{Q_1}$,
$q = q_1 + \tilde{q}_1$. Introducing a single field $\phi$ of $U(1)_X$ charge
normalized to -1, it is straightforward to construct
a supersymmetric and gauge invariant Lagrangian. It will
be completely determined by the gauge kinetic functions

\begin{equation}
f_a \ = \ \frac{T}{2 \pi},
\label{fa}
\end{equation}

\noindent
the Kahler potential

\begin{equation}
K \ = \ - \ 3 \ \ln [ T + \overline{T} - |\phi|^2 - |\chi|^2 ] \ ,
\label{K}
\end{equation}

\noindent
and the gauge invariant superpotential

\begin{equation}
W \ = \ W_0  \ + \ \frac{c}{\chi^2} \ e^{-aT} \ + \ \tilde{m} \phi^q\chi^2 \ . 
\label{W}
\end{equation}

\noindent
Gauge invariance of the nonperturbative term fixes the mesons charges to be 
\begin{equation}
q \ = \ {a \over 4} \ \delta_{GS} \ . \label{gaugeinv}
\end{equation}
We can then compute the F-term  and D-term contributions $V_F$ and $V_D$ to the scalar potential

\begin{eqnarray}
&& V_F \ = \ e^K \ \left( K^{I {\bar J}} D W_I {\bar D} {\bar W}_{\bar J} \ - \ 3 |W|^2  \right)  \nonumber \\  
&& = \ \frac{1}{r^3} \{
\frac{r^2}{3}|\partial_T W - \frac{3}{r} W|^2
+\frac{r}{3} \sum_{i=1}^{2}
|\partial_i W + \overline{\phi_i} \partial_T W|^2
-3 |W|^2  
\} \ , \label{VF}
\end{eqnarray}

\begin{equation}
V_D \ = \ \frac{4 \pi}{T + \overline{T}}
\left(
q_i \phi^i  \partial_i K  \ - \ {\delta_{GS} \over 2} \ \partial_T K \right)^2 \ = \ 
\frac{36 \pi}{(T + \overline{T}) r^2 } \left[ -|\phi|^2 +\frac{q}{2}|\chi|^2+ {\tilde \xi}^2 \right]^2 \ , 
\label{VD}
\end{equation}
\noindent where ${\tilde \xi}^2 = {3 \delta_{GS} / 2} > 0$, $\phi_i= \phi, \chi$ and we have introduced
$r=(T + \overline{T} - \sum_i|\phi_i|^2)$. Substituting the
Kahler potential and superpotential given by Eqs.(\ref{K}) and
(\ref{W}), we can write the total potential as

\begin{equation}
\begin{array}{c}
V_{tot} \ = \ V_F \ + \ V_D \ = \ 
{1 \over 3 r} |\frac{ace^{-aT}}{\chi^2}+\frac{3}{r}W|^2
+\frac{1}{3r^2}|q\tilde m \phi^{q-1}\chi^2-
\overline{\phi}\frac{ace^{-aT}}{\chi^2}|^2
\\
+\frac{1}{3r^2}|\frac{-2ce^{-aT}}{\chi^3}+2 \tilde m \phi^q\chi
-\overline{\chi}\frac{ace^{-aT}}{\chi^2}|^2
-\frac{3}{r^3}|W|^2+\frac{36 \pi}{r^2(T+\overline{T})}
\left[ -|\phi|^2 +\frac{q}{2}|\chi|^2+ {\tilde \xi}^2 \right]^2 \ . 
\end{array}
\label{Vtot}
\end{equation}

\subsection{The solution of the equation of motions}

\noindent
The minimum of the potential (\ref{Vtot}) is obtained by making
an expansion around the fields configuration
($T_0$, $\phi_0$, $\chi_0$)

\begin{eqnarray}
&&\phi_{min} \ = \ \phi_0 \ (1 + \epsilon \phi_1 + \epsilon^2 \phi_2) \ , \nonumber \\
&& \chi_{min} \ = \ \chi_0 \ (1 + \epsilon \chi_1) \ , \nonumber \\
&& T_{min} \ = \ T_0 \ , \label{sol1}
\end{eqnarray}

\noindent
where $\epsilon$ is the parameter of the expansion, proportional
to the ratio of the dynamic scale to the fundamental scale
($M_0/ {\tilde \xi}^2$ in the notation of this section). In (\ref{sol1}), $\phi_0$ and $\chi_0$ are 
the solutions of the Eqs. of motion by considering the two relevant scales, $M_0$ and ${\tilde \xi}^2$, decoupled or,
equivalently, infinitely far apart and also neglecting the supergravity corrections to the scalar potential. 
Consequently, they are defined by the equations of the global supersymmetry D and F flatness
\begin{eqnarray}
&& D^{(0)} \ = \ - |\phi_0|^2 + {\tilde \xi}^2 \ = \ 0 \ , \nonumber \\
&& F_{\chi}^{(0)} \ = \  - {2 c \over \chi_0^3} e^{-a T_0} \ + 2 {\tilde m} \phi_0^q \chi_0 \ = \ 0 \ . \label{sol2}
\end{eqnarray} 
The explicit value of $T_0$ is not needed in finding $\phi_{min}$ and $\chi_{min}$ and will be specified later on, 
but for consistency 
of the supergravity analysis it has to satisfy $T_0  \gg 1$ in supergravity units. 
At first sight, it can seem strange to develop $\phi$ at second order
into $\epsilon$ and $\chi$ only at first order. In fact, as we will
see later, we need to calculate $\phi$ up to second order in order to get the leading D-term
contribution. $T$ can be evaluated at  lowest order in the $\epsilon$ expansion, as any higher order
corrections does no contribute significantly to the vacuum energy. 
In the limit of interest $M_0 \ll {\tilde \xi}^2 \ll M_P^2$, the leading terms in the field equations for $\phi$ and $\chi$ 
are the ones coming from the global supersymmetry limit 
\begin{equation}
{\partial V \over \partial \phi} \simeq {\partial V_{\rm global} \over \partial \phi} \ = \ 0 \quad , \quad 
{\partial V \over \partial \chi} \simeq {\partial V_{\rm global} \over \partial \chi} \ = \ 0 \  \label{sol3}
\end{equation}
and they are solved precisely as in \cite{Binetruy:1996uv} in a power expansion in $\epsilon$. The minimization with respect to
$T$,
\begin{equation}
{\partial V \over \partial T}  \ = \ 0 \label{sol4}
\end{equation} 
has clearly to be done with the full SUGRA potential (\ref{Vtot}). The intuitively natural zeroth order ansatz 
$F_T^{(0)}  = 0$, which would imply 
$W_0 \sim {\tilde m} \phi_0^q \chi_0^2$ turns out to be wrong, i.e. not to lead to a solution of (\ref{sol4}). 
The self-consistent solution, in our approximations, turns out to be obtained by
\begin{equation}
W_0 \ \simeq \ - { {\tilde m} q^2 \over 3 a} \ \phi_0^{q-2} \ \chi_0^2 \ . \label{sol5} 
\end{equation}
Finally, the complete set of solutions, including the two subleading terms in $\phi$ and one subleading term in $\chi$, needed
for a consistent treatment, are given by 
\begin{equation}
\phi_{min} \ = \ {\tilde \xi}
\left(1-\frac{3}{4}\frac{a W_0}{\tilde m q {\tilde \xi}^q}
-\frac{9}{32}\frac{a^2 W_0^2}{q^2 {\tilde m}^2 {\tilde \xi}^{2q}}
-\frac{9}{16}\frac{a^2 W_0^2}{{\tilde m}^2 q {\tilde \xi}^{2q}}
+\frac{(q-2) a W_0^2}{24 \pi q^2 {\tilde \xi}^2}
\ln \left[
\frac{9 a^2 W_0^2}{c \tilde m q^4 {\tilde \xi}^{q-4}}
\right] \right) \label{phimin}
\end{equation}

\begin{equation}
\chi_{min} \ = \ \sqrt{\frac{-3 a W_0}{\tilde m q^2 {\tilde \xi}^{q-2}}}
\left( 1 \ + \ \frac{3}{8}\frac{a W_0}{\tilde m {\tilde \xi}^q}
\right) \ , \label{chimin}
\end{equation}

\begin{equation}
T_{min} \ = \ - \ \frac{1}{a} \ \ln \left[
\frac{9 a^2 W_0^2}{c \tilde m q^4 {\tilde \xi}^{q-4}}
\right] \ . \label{tmin}
\end{equation}

Notice that the vev of $\phi$ generates effectively a mass for the mesons $m_{\chi} \sim {\tilde m} 
{\tilde \xi}^q$. 
It is interesting to point out here two kind of approximation we'll
use to find the coefficients $\epsilon \phi_1$, $\epsilon^2 \phi_2$
and $\epsilon \chi_1$ : $\chi_0 \ll \phi_0 \ll M_P$.
The first approximation is coming from the global SUSY ansatz
$F_{\chi} \sim \partial_{\chi} W = 0$. The second one is necessary for having an effective supergravity description.
The expansion parameter is then given by
\begin{equation}
\epsilon \ = \ {M_0 \over {\tilde \xi}^2} \ = \ {-3 a W_0 \over 2 q^2 {\tilde m} {\tilde \xi}^q} \ = \
- \ {-3 a W_0 \over q^2 m_{\chi} M_P^2} \ \ll 1 \ , \label{epsilon}  
\end{equation}
where in the list equality we reinstalled the Planck mass. From the solutions (\ref{phimin})- (\ref{tmin}) , 
we can easily extract the values of the different
auxiliary fields $F^{\phi}$, $F^{\chi}$ and $F^{T}$ at the minimum
of the potential :

\begin{eqnarray}
&& F^{\phi} \ = \ \frac{W_0 {\tilde \xi}}{\sqrt{2 T_0}}
\left[ \frac{a}{q} + \frac{1}{r} -
\left(
\frac{a {\tilde \xi}}{q}
\right)^2 \right] \ , \nonumber \\
&& F^{\chi} \ = \ \frac{W_0}{\sqrt{2 T_0}}
\sqrt{\frac{-3 a W_0}{\tilde m q^2 {\tilde \xi}^{q-2}}}
\left[ -\frac{5}{8} \frac{a}{\tilde m {\tilde \xi}^q}
+\frac{1}{r}
-\left(
\frac{a {\tilde \xi}}{q}
\right)^2
\right] \ , \nonumber \\
&& F^T \ = \ \frac{W_0}{\sqrt{2 T_0}}
\left[ 2 T_0
\left( \frac{1}{r}
-\left( \frac{a {\tilde \xi}}{q}
\right)^2 \right)
+\frac{a {\tilde \xi}^2}{q^2}
\right] \ , \nonumber \\
&& D \ = \ \frac{3(q-2)a^2 W_0^2}{24 \pi q^2} \ , \label{Dterm}
\end{eqnarray}
where we defined $F^i = K^{i {\bar j}} \exp (K/2) D_{\bar j} {\bar W}$. 
From the auxiliary fields, it becomes straightforward to calculate
the vacuum energy at the minimum :

\begin{equation}
V_0 \ \simeq \ - \ \frac{3 a^2 W_0^2 {\tilde \xi}^2}{4 q^2 T_0^2} \ 
 + \ \frac{(q-2)^2 a^4 W_0^4}{32 \pi T_0 q^4} \ , \label{V0}
\end{equation}

where the first, negative term is the F-term contribution, while the second, positive one is the D-term contribution. 
It is interesting to calculate the value we need for $W_0$ in
order to have a zero cosmological constant. We get

\begin{equation}
W_0^{\mathrm{zero ~ cc}} \ \simeq \ \sqrt{\frac{24 \pi}{T_0}}
\frac{{\tilde \xi} q}{a(q-2)} \ .  \label{W0null}
\end{equation}

\noindent
It becomes clear from Eq.(\ref{W0null}) that the FI scale is
tightly linked with the gravitino mass (given roughly by
$W_0$ in Planck scale units). In another words
a very small FI scale is needed to achieve a TeV SUSY
spectrum. By using (\ref{gaugeinv}) it turns out that the gravitino mass and the FI term are related by $m_{3/2} 
\sim W_0/M_P^2 \sim {\tilde \xi}^4 /M_P^3$,
so TeV scale gravitino would imply roughly ${\tilde \xi} \sim 10^{15} GeV$. Whereas this is maybe not imposible to imagine 
in certain orientifold models,
(\ref{gaugeinv}) then implies unnaturally small meson charges $q \sim 10^{-8}$. 
TeV scale gravitino mass is therefore very difficult to obtain. This
may be a phenomenological shortcoming if one attempts to get
conventional low--energy SUSY. A similar result were obtained by the authors of
\cite{Achucarro:2006zf}, which did use non--massive mesons fields.

Our results can be understood qualitatively by noticing that, by defining the gravitino mass
\begin{equation}
m_{3/2} \ = \ W e^{K/2} \ \simeq \ {W_0 \over 2{\sqrt 2} T_0^{3/2}} \ , \label{sol6} 
\end{equation}
at the minimum the F-terms and the D-term are of the order (we neglect here numerical factors)
\begin{eqnarray}  
&& {F^{\phi} \over \phi_0} \ \sim \ {F^{\chi} \over \chi_0} \ \sim \ F^T \ \sim \ T_0 \ m_{3/2} \ , \nonumber \\
&& D \ \sim \ T_0^3 m_{3/2}^2 \ .  \label{sol7}
\end{eqnarray}
Consequently, the F and D contributions to the vacuum energy in (\ref{V0}) are qualitatively of the form $\langle V \rangle = V_F + V_D \sim - 
 m_{3/2}^2 {\tilde \xi}^2 +  m_{3/2}^4 $ and their cancellation can only occurs for a very large gravitino mass, modulo 
the case of very small FI term discussed in the previous paragraph.   

Finally, we would like to point out that the analysis in this section and the next one, although described
in terms of the Type IIB set--up used by KKLT, is valid with small changes also for models coming from heterotic
string construction.

\subsection{Stabilization with positive vacuum energy}

We have seen that positive FI term does not give a sufficient
contribution to the D-term scalar potential to obtain a zero
cosmological constant. However, it was recently stressed by
the authors of \cite{Dudas:2005vv}
that negative FI terms could lead to important D--terms in the case $q=1$, i.e. the meson masses come from a Yukawa type
term in the superpotential $\phi \chi^2$. Due to the non universal nature of the anomaly cancellation conditions
in orientifold models, this is a new possibility to explore.  
It is natural to see if it is possible to stabilize the moduli fields
in this case. The microscopic model we have in mind for this section is the following. Consider an intersecting brane model
like in section 2.2, with fluxes on a stack of branes containing both the anomalous $U(1)$ and the hidden sector gauge group,
such that $m_2,m_3 > 0$, whereas $m_1  < 0$. In this case, the modulus $T_1$ appears with a negative coefficient in the gauge
kinetic function (\ref{gi10}) and also in the induced FI term (\ref{gi12}). If the other two Kahler moduli $T_2,T_3$ are
stabilized in a supersymmetric way, then the only unstabilized modulus $T_1$ realizes the case ii) in the 
section 5 of \cite{Dudas:2005vv} where the sign of the FI term is such that the fields which have to condense
in order to compensate it are the composite, hidden sector mesons. As we already mentioned in section 2.1, this
is impossible to realize in the heterotic case, but possible in orientifold models with  magnetized D-branes. 
Notice that generically the resulting $U(1)$ gauge symmetry becomes a gauged R-symmetry, however in the special case
where after stabilization the resulting FI term has no pure constant (i.e. $S_0,T_2,T_3$ dependent ) piece, it is still
a regular gauge symmetry. As will become clear from our discussion below, this is more interesting case to analyze.
If the charge $q=1$, according to \cite{Dudas:2005vv} we expect a large
D-term. The effective action in this situation is
\begin{eqnarray}
&& K \ = \ - \ \ln (T_1 + {\bar T}_1 - \sum_i |\phi|^2) \quad , \quad 
f_h \ = \ (\prod_{i=1}^3 n_i^h) S_0 - \alpha_h T_1 \ , \nonumber \\
&& f_{U(1)} \ = \  (\prod_{i=1}^3 n_i^{U(1)}) S_0 - \alpha_1 T_1 \quad , \quad 
W \ = \ W_0 \ + \ {c \over \chi^2} e^{\alpha_h T_1} + \lambda \phi \chi^2 \ , \label{neg01} 
\end{eqnarray}
where $n_i^h$ ($n_i^{U(1)}$) are here the wrappping numbers for the hidden sector branes ($U(1)$ brane),
where $\phi_i = \phi, \chi$, $c$ is an effective (exponentially small ) constant $c \sim \exp (- |n_1^h  n_2^h  n_3^h|  
S_0 )$ and $S_0$ is the stabilized value of the dilaton.   
The F and D-term contributions to the scalar potential in this case are found to be
\begin{eqnarray}
&& V_F \ = \  \ \frac{1}{r} \{
r^2 |\partial_T W - \frac{1}{r} W|^2
\ + \ r \sum_i |\partial_i W + \overline{\phi_i} \partial_T W|^2
-3 |W|^2 \}  \ \nonumber \\
&& =  r \ |\frac{\alpha_h c e^{\alpha_h T_1}}{\chi^2} \ - \ \frac{1}{r}W|^2
\ +\ |  \lambda \chi^2 \ + \
\overline{\phi} \ \frac{\alpha_h c e^{\alpha_h T_1}}{\chi^2}|^2
\nonumber \\
&& + |\frac{-2 c \ e^{\alpha_h T_1}}{\chi^3} \ + \ 2 \lambda \phi \chi
 +  \overline{\chi} \ \frac{\alpha_h c e^{\alpha_h T_1}}{\chi^2}|^2
-\frac{3}{r} |W|^2 \ , \nonumber \\   
&& V_D \ = \ \frac{4 \pi}{r^2 [S_0 + {\bar S}_0- \alpha_1 (T_1+\overline{T}_1)]}
\left[ -|\phi|^2 +\frac{1}{2}|\chi|^2 -  \xi'^2 \right]^2 \ , 
\label{VDchineg}
\end{eqnarray}
where $r \ = \ T_1 + {\bar T}_1 - \sum_i |\phi|^2 $. 
\noindent
The D term contribution turns out to be large by minimization due to the superpotential term $W = \lambda \ \phi
\ \chi^2$ which forbids the field $\chi$ to efficiently cancel the large and negative FI term. Notice, as already emphasized, that in this case the 
composite meson field $\chi$ has the correct charge to cancel the FI term. 
For a fixed $T=T_{min}$, we find minima for the other two fields 
\begin{eqnarray}
&& \phi_{min} \ \simeq \ \frac{c \ e^{\alpha_h T_{min}}}{4 \lambda \xi'^4}
\left[ 1 + {\lambda^2 [S_0 + {\bar S}_0- \alpha_1 (T+\overline{T})] r^2 \over  \pi}  
\right]^2  \ , \nonumber \\
&& \chi_{min} \ \simeq \ \sqrt{2} \ \xi' \left[ 1 + {\lambda^2 [S_0 + {\bar S}_0- \alpha_1 (T+\overline{T})] r^2 \over \pi}  
\right]^{-1/2}  \ . \label{neg1}
\end{eqnarray}
Stabilization with respect to $T$ then gives the minimization equation
\begin{equation}
W_0^2 \ \simeq \ {2 \lambda^4 r^3 \over \pi} [ 2( S_0 + {\bar S}_0) - 3 \alpha_1 (T +{\bar T})]  \left[ 1 + 
{\lambda^2 [S_0 + {\bar S}_0- \alpha_1 (T+\overline{T})] r^2 \over  \pi}  
\right]^{-2} \ \xi'^4 \ . \label{neg02} 
\end{equation}
At the minimum, the auxiliary fields are of the order
\begin{equation}
F^{\phi} \ \sim \ \ \lambda \ \xi'^2 \quad , \quad F^{\chi} \ \sim \  c {e^{\alpha_h T_{min}} \over \xi'^4} \quad , \quad
D \ \sim \ \lambda \ \xi'^2 \ . \label{neg2}
\end{equation}
In order to be able to compensate the large positive vacuum energy from the D-term, we need in this case $W_0 >> (c e^{\alpha_h T_{min}} / \chi_0^2)$,
in which case the vacuum energy cancellation reads approximatively
\begin{equation}
{2 \over r} |W_0|^2 \ \simeq \ \lambda^2 |\chi_0|^4 \ + \ {{\tilde g}^2 \over 2} D^2 \ , \label{neg03} 
\end{equation}
where ${\tilde g}^2 = (4 \pi)/[(S_0-\alpha_1 T_1) r^2]$. Combining (\ref{neg02}) and (\ref{neg03}), we find
\begin{equation}
T_{min}^2 ( S_0- 2 \alpha_1 T_{min} )  \ \simeq \ {\pi \over 8 \lambda^2} \ . \label{neg04}
\end{equation}
We therefore need a relatively small $\lambda$ in order to get a relatively large $T_1$. 
From these solutions, we can easily understand why the
contribution from $V_D$ is large. Indeed, whereas
$V_D \propto \mathrm{(dynamical ~~ scale)^8}$ when the FI term is
positive (giving a negligible contribution to the
cosmological constant),
whereas $V_D \propto \mathrm{(FI ~~ scale)^4}$ for a negative FI term.


At the minimum, the FI term and the gravitino mass are related qualitatively as $\xi'^2 \sim m_{3/2} M_P / \lambda $. Natural
values for FI term $\xi \sim 10^{-2} M_P$ and $\lambda \sim 10^{-2}$ then lead to $m_{3/2} \sim 10^{-6} M_P$. This is an
improvement compared to the case of (more traditional) positive FI term of the last section, but is still far from
the TeV range.  
Gravitino mass in the TeV range is in principle possible for intermediate (of the order $10^{12}$ GeV ) values of FI term. 
This looks again like a fine-tuning, though internal magnetic fields could maybe dynamically produce
such anomalously low values. 

As mentioned at the beginning of this section, stabilization of the other Kahler moduli and of the dilaton, which tranform 
non-linearly under the $U(1)$ gauge transformations, generically lead to a $U(1)$ gauged R-symmetry. In this case,
the constant $W_0$ in (\ref{neg01}) is not allowed. If $W_0$ is absent, a similar analysis as before shows that the vacuum
energy will always be positive. Alternatively, $W_0$ could be replaced eventually with some dynamically generated
term of appropriate R-charge $W_0 \sim \phi^x$. We didn't investigate the outcome of this last possibility.

Our conclusion for this section is that with (negative) FI terms partially compensated by composite hidden sector 
fields, it is generically easier to get zero or positive vacuum energy, even if the natural value of the gravitino 
mass is of intermediate scale values instead of TeV scale. These class of models deserve therefore, in our opinion,
more detailed theoretical and phenomenological studies.   

\section{Kahler uplifting}

Previous works in the context of heterotic effective supergravity \cite{Binetruy:1996xj,
Casas:1996zi} and more recent one \cite{Dudas:2005vn,bhk}  underlined the role played by
corrections to the Kahler metric for the moduli stabilization with a zero or positive vacuum energy.
It is natural to study then their implications on the scalar
potential uplifting, and to try to find conditions to obtain a
consistent model with a dS vacuum and a TeV-scale supersymmetric
spectrum. We will consider a KKLT--like scenario with a
superpotential

\begin{equation}
W \ = \ W_0 \ + \ c e^{-aT},
\end{equation}

\noindent
with $W_0$ determined by the flux compactification process \footnote{By simplicity, we will consider in our discussion
$W_0>0$ and $c>0$ whereas the authors of \cite{Kachru:2003aw} took $W_0<0$.
This is just a phase definition that can be absorbed by the axion
field as it will be clear later on.}.
However instead of adding the effect of a non--supersymmetric
$\overline{D3}$--brane or D-term contributions generated by magnetic fluxes, we consider perturbative
corrections to the Kahler metric of the general form
\footnote{In what follows we will use units where $M_P = 1$.} :

\begin{equation}
K \ = \ - \ 3 \ \ln \ [T+\overline{T}] \ - \ \frac{\Delta (T, {\bar T})}{T + \overline{T}} \ , \label{kahler1}
\end{equation}

\noindent
$\Delta$ being a real function of the $T$ field
$\Delta = f( T + \overline{T})$ which we consider for consistency reasons to be small $\Delta \ll 1$ in the large volume
limit $T \ll 1$. It is straightforward to compute
the scalar potential, keeping the dominant terms in $\Delta$ in the expansion :

\begin{equation}
V(T) \ \sim \ \frac{1}{(T + \overline{T})^2}
\left[
c a (e^{-aT} W_0 \ + \ e^{-a \overline{T}} \overline{W_0})
\ + \ |W_0|^2 \ \Delta''
\right] \ . \label{VknpT}
\end{equation}

\noindent
Notice that in order for Eq. (\ref{VknpT}) to be valid, we must have
$|c a e^{-aT}| \ll |W_0|$ (considering $\Delta$ as a perturbation to the
Kahler metric). Developing $T = t + i \theta$ and considering
$W_0$ as a real parameter, (\ref{VknpT}) can be re--expressed as

\begin{equation}
V(t,\theta) \ = \ \frac{1}{4 t^2}
\left[
2 c a W_0 e^{-at} \ \cos(\theta a)
\ + \ |W_0|^2 \ \Delta''(t)
\right] \ . \label{Vknptheta}
\end{equation}

The minimization of (\ref{Vknptheta}) is straightforward and leads to

\begin{equation}
\theta_0 = \pi / a ~~~~~~~~~~~ {\rm modulo} \ [2 \pi/ a] \ ,
\end{equation}

\noindent
giving for $V(t)$

\begin{equation}
V(t) \ = \ V(t,\theta_0) \ = \ \frac{1}{4 t^2}
\left[ -2 c a W_0 e^{-at} \ + \ |W_0|^2 \ \Delta''(t)
\right] \ .
\label{Vknpthetamin}
\end{equation}

We understand here why the relative sign between $W_0$ and $c$ is
just a convention as the phase can be reabsorbed by the axion field
$\theta$ at the minimum. Notice also that the non--perturbative effects
are fundamental for the stabilization of the axion. Indeed, the corrections
to the Kahler metric are blind to the axion field which can only
appear in the scalar potential through the superpotential. In
other words, a model with only "pure" Kahler contributions is not
able to stabilize $\theta$.

Form (\ref{Vknpthetamin}) we can deduce the two conditions we need
to have a Minkowski solution at the minimum ($T_0=t_0+i\theta_0$) :

\begin{equation}
\frac{\partial V}{\partial t}|_{t_0} \ = \ 0
~~~~~~\Rightarrow
~~~~~~\Delta'''|_{t_0} \ = \ - \ \frac{2 \ c \ a^2 \ e^{-a t_0}}{W_0}
\label{minim}
\end{equation}

\noindent
and

\begin{equation}
V(t_0) = 0
~~~~~~\Rightarrow
~~~~~~\Delta''|_{t_0} \ = \ \frac{2 \ c \ a \ e^{-a t_0}}{W_0}.
\label{Minkows}
\end{equation}

\noindent
Combining (\ref{minim}) and (\ref{Minkows}) we obtain

\begin{equation}
\frac{\Delta'''}{\Delta''}|_{t_0} \ = \ - a \ , 
\end{equation}

showing the connexion between non--perturbative
effects and the corrections to the Kahler potential.

From (\ref{Vknpthetamin}), we can deduce some general arguments
concerning the function $\Delta(t)$ in order to satisfy our basic constraints.
Indeed, the potential must vanish positively when $t \rightarrow \infty$.
It implies that the positive contribution, proportional to $\Delta''(t)$,
must decrease slower than the non--perturbative contribution
(proportional to $e^{-a t}$). It means that $\Delta(t)$ can be constructed
by the combination of polynomial forms ($\alpha/t^\gamma$) and
exponential functions
($\alpha' e^{-\beta t}/t^{\gamma'}$ with $\beta < a$).

It is interesting to calculate the mass scales of such a model.
The gravitino mass, $m_{3/2}=e^{K/2} W$, auxiliary field $F^t$,
moduli and axion masses $m_t$, $m_{\theta}$ are at the minimum
$t_0$ :

\begin{eqnarray}
&&
m_{3/2} \ \sim \ \frac{W_0}{(2t_0)^{3/2}} \quad , \quad  F^T \ \sim \ \frac{W_0}{(2t_0)^{1/2}}
 \ , \nonumber \\
&&
m_t^2 \ \sim \ \frac{1}{6}
\left[
-2 \ c \ a^3 \ e^{-a t_0} \ W_0 \ + \ W_0^2 \ \Delta^{(4)}(t_0)
\right] \ , \nonumber \\
&&
m_{\theta}^2 \ = \ \frac{1}{3} \ c \ a^3 \ W_0 \ e^{-a t_0} \ . 
\end{eqnarray}

\noindent
Contrary to the pattern of mass of the KKLT scenario
studied in \cite{Choi:2004sx} where a "little hierarchy" appeared
between the SUSY spectrum ($\sim F^T/T$) and the gravitino mass,
we don't observe such difference in our model. Moreover,
we obtain lighter moduli fields than in KKLT scenario :
$m_t^2$ is suppressed by a factor $\Delta'' \times t_0^3$ with respect to
gravitino mass squared. It would be very interesting to study the stability of the vacua, along the lines of \cite{scrucca}, 
in the presence of generic corrections to the Kahler potential.  

\subsection{An example}

To illustrate , we apply our discussion to the phenomenological model developed
in \cite{Dudas:2005vn}, where $\Delta$ was computed from the one-loop vacuum (Casimir) energy with massless and massive
fields in the higher-dimensional bulk space. In \cite{Dudas:2005vn} supersymmetry was broken by boundary conditions and
consequently a superpotential $W = \omega$ and a correction to the Kahler potential $\Delta (\omega, T, {\bar T})$ were generated.
The first step in our present investigation is the supersymmetric limit $\omega \rightarrow 0$, in which 
$\Delta$ has a well-defined limit, whereas the superpotential vanishes. This is consistent with the fact that the superpotential
is non-renormalized under quantum corrections, whereas the Kahler potential is renormalized. The second step is to add
the KKLT type superpotential generated by fluxes and gaugino condensation on
 $D7$-branes, which is unchanged by the
the one-loop quantum corrections. 

\begin{figure}
    \begin{center}
\centerline{
       \epsfig{file=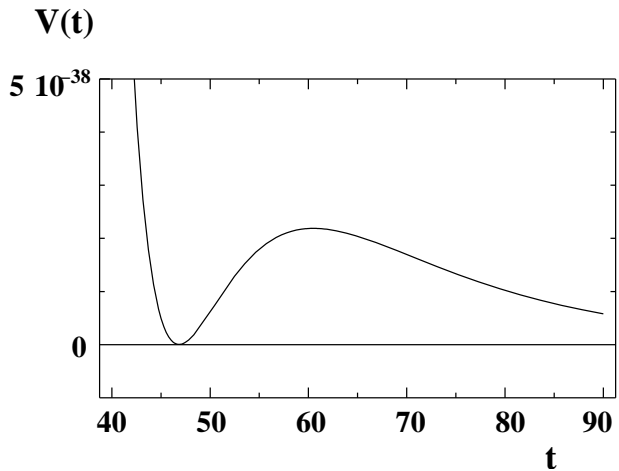,width=0.5\textwidth}
       }
          \caption{{\footnotesize
Scalar potential with correction to the Kahler potential
of the form (\ref{DeltaEmilian}) for
for $W_0=10^{-13}$, $c=10$, $a=4$; $\alpha_0=3.824$,
$\alpha_1=10$, $\beta_1=0.09$;
$\alpha_2=-50$ and $\beta_2=0.04$.
}}
\label{fig:Vknp}
    \end{center}
\end{figure}

In this case, $ \Delta $ is given by :

\begin{equation}
\Delta (T, {\bar T})  \ = \ \frac{1}{(T + \overline{T})^2}
\left[ \alpha_0 + \alpha_1 (T+ \overline{T}) e^{-\beta_1 (T+\overline{T})}
+\alpha_2 e^{-\beta_2 (T+\overline{T})}
\right] \ , \label{DeltaEmilian}
\end{equation}
where the exponential terms come from contributions to the vacuum energy \\
$\exp [-m (T+{\bar T})]$ from massive fields of mass $m$.  
The coefficients $\alpha_i, \beta_i$ in (\ref{DeltaEmilian}) are the $\omega \rightarrow 0$ limit of the values displayed
in \cite{Dudas:2005vn}.
To stabilize T with a dS vacuum and an acceptable
behavior of the potential for $T \sim 0$ and large value of $T$,
it is easy to check that we need $\alpha_0 > 0$, $\alpha_1 > 0$ and
$\alpha_2 < 0$, whereas, as discussed above, $\beta_1$, $\beta_2 <a$.
We illustrated the scalar potential for specific values
of the parameters ($c,~a,~\alpha_i,~\beta_i$)
in Fig. (\ref{fig:Vknp}), which clearly illustrate that TeV values for the gravitino mass do not require extremely
small values of the fundamental parameters in the Kahler potential.  
In \cite{bhk} explicit tree-level and one-loop string corections to the Kahler potential were included in the discussion
of moduli stabilization. In the large volume limit, these corrections are probably bigger than the corrections we kept
in our example (\ref{DeltaEmilian}). We believe that the Kahler corrections invoked in
\cite{bhk}, combined with exponential corrections $\exp (-\beta (T + {\bar T}))$ in $\Delta$ coming from massive bulk 
fields produce a stabilization pattern similar to the example we displayed in this section. 

\section*{Acknowledgments}{ Work partially supported by the CNRS PICS \#~2530
and 3059, RTN contracts MRTN-CT-2004-005104 and MRTN-CT-2004-503369,  the European Union Excellence Grant, MEXT-CT-2003-509661.
The work of Y.M. is sponsored by the PAI programm PICASSO under 
contract PAI--10825VF.
The authors want to thank P. Binetruy, Z. Lalak, S. Pokorski and S. Vempati for very useful discussions. E.D. 
thanks the Galileo Institute of Theoretical Physics and INFN
for partial support, while Y.M. would like to acknowledge the hospitality of the CERN Theory
Division during the completion of this work.  }


\nocite{}
\bibliography{bmn}
\bibliographystyle{unsrt}

\end{document}